\documentclass[%
 reprint,
 superscriptaddress,
 nofootinbib,
 amsmath,amssymb,
 aps,prc
]{revtex4-2}

\usepackage{graphicx}
\usepackage{dcolumn}
\usepackage{bm}\usepackage{siunitx}
\usepackage{multirow}
\usepackage{braket}
\usepackage[final,hidelinks]{hyperref}

\begin{document}

\preprint{APS/123-QED}

\title{Improved measurement of the \texorpdfstring{$0_2^+\rightarrow0_1^+$}{0\_2->0\_1} E0 transition strength for \texorpdfstring{$^{72}$Se}{72Se} using the SPICE spectrometer}

\author{J. Smallcombe}
\email[Corresponding Author: ]{smallcombe.james@jaea.go.jp}
\affiliation{Advanced Science Research Center, Japan Atomic Energy Agency (JAEA), Tokai, Ibaraki 319-1195, Japan}
\affiliation{Oliver Lodge Laboratory, University of Liverpool, Liverpool L69 9ZE, United Kingdom}
\author{A.B. Garnsworthy}
\affiliation{TRIUMF, 4004 Wesbrook Mall, Vancouver, British Columbia, Canada, V6T 2A3}

\author{W. Korten}
\affiliation{Irfu, CEA, Universit\'e Paris-Saclay, F-91191 Gif-sur-Yvette, France}

\author{P. Singh}
\affiliation{Irfu, CEA, Universit\'e Paris-Saclay, F-91191 Gif-sur-Yvette, France}

\author{F.A. Ali}
\affiliation{Department of Physics, University of Guelph, Guelph, ON, N1G 2W1, Canada}
\affiliation{Department of Physics, University of Sulaimani, Kurdistan Region, Iraq}

\author{C. Andreoiu}
\affiliation{Department of Chemistry, Simon Fraser University, Burnaby, British Colomia V5A 1S6, Canada}

\author{S. Ansari}
\affiliation{Irfu, CEA, Universit\'e Paris-Saclay, F-91191 Gif-sur-Yvette, France}

\author{G.C. Ball}
\affiliation{TRIUMF, 4004 Wesbrook Mall, Vancouver, British Columbia, Canada, V6T 2A3}

\author{C.J. Barton}
\affiliation{Department of Physics, University of York, Heslington, York YO10~5DD, United Kingdom}

\author{S.S. Bhattacharjee}
\altaffiliation[Present address: ]{Institute of Experimental and Applied Physics, Czech Technical University in Prague, Husova 240/5, 110 00, Prague 1, Czech Republic}
\affiliation{TRIUMF, 4004 Wesbrook Mall, Vancouver, British Columbia, Canada, V6T 2A3}

\author{M. Bowry}
\altaffiliation[Present address: ]{School of Engineering, Computing and Physical Sciences, University of the West of Scotland, High Street, Paisley PA1 2BE, United Kingdom}
\affiliation{TRIUMF, 4004 Wesbrook Mall, Vancouver, British Columbia, Canada, V6T 2A3}

\author{R. Caballero-Folch}
\affiliation{TRIUMF, 4004 Wesbrook Mall, Vancouver, British Columbia, Canada, V6T 2A3}

\author{A. Chester}
\altaffiliation[Present address: ]{National Superconducting Cyclotron Laboratory, Michigan State University, East Lansing, Michigan 48824, USA}
\affiliation{TRIUMF, 4004 Wesbrook Mall, Vancouver, British Columbia, Canada, V6T 2A3}

\author{S.A. Gillespie}
\altaffiliation[Present address: ]{National Superconducting Cyclotron Laboratory, Michigan State University, East Lansing, Michigan 48824, USA}
\affiliation{TRIUMF, 4004 Wesbrook Mall, Vancouver, British Columbia, Canada, V6T 2A3}

\author{G.F. Grinyer}
\affiliation{Department of Physics, University of Regina, Regina, Saskatchewan, S4S 0A2, Canada}

\author{G. Hackman}
\affiliation{TRIUMF, 4004 Wesbrook Mall, Vancouver, British Columbia, Canada, V6T 2A3}

\author{C. Jones}
\altaffiliation[Present address: ]{School of Computing, Engineering and Mathematics, University of Brighton, Brighton BN2 4GJ, United Kingdom}
\affiliation{Department of Physics, University of Surrey, Guildford, Surrey, GU2 7XH, United Kingdom}

\author{B. Melon}
\affiliation{INFN Sezione di Firenze I-50019, Italy}

\author{M. Moukaddam}
\altaffiliation[Present address: ]{Universit\`e de Strasbourg, IPHC, 23 rue du Loess, 67037 Strasbourg, France}
\affiliation{Department of Physics, University of Surrey, Guildford, Surrey, GU2 7XH, United Kingdom}

\author{A. Nannini}
\affiliation{INFN Sezione di Firenze I-50019, Italy}

\author{P. Ruotsalainen}
\affiliation{University of Jyv\"askyl\"a, Department of Physics, P.O. Box 35, FI-40014 University of Jyv\"askyl\"a, Finland}

\author{K. Starosta}
\affiliation{Department of Chemistry, Simon Fraser University, Burnaby, British Colomia V5A 1S6, Canada}

\author{C.E. Svensson}
\affiliation{Department of Physics, University of Guelph, Guelph, ON, N1G 2W1, Canada}

\author{R. Wadsworth}
\affiliation{Department of Physics, University of York, Heslington, York YO10~5DD, United Kingdom}

\author{J. Williams}
\affiliation{Department of Chemistry, Simon Fraser University, Burnaby, British Colomia V5A 1S6, Canada}

\date{\today}

\begin{abstract}
The selenium isotopes lie at the heart of a tumultuous region of the nuclear chart where shape coexistence effects grapple with neutron-proton pairing correlations, triaxiality, and the impending proton dripline. In this work a study of $^{72}$Se by internal conversion electron and $\gamma$-ray spectroscopy was undertaken with the SPICE and TIGRESS arrays. New measurements of the branching ratio and lifetime of the $0_2^+$ state were performed yielding a determination of $\rho^2(E0;0_2^+{\rightarrow}0_1^+)=29(3)$ milliunits.
two state mixing calculations were performed that highlighted the importance of interpretation of such $E0$ strength values in the context of shape-coexistence.
\end{abstract}

\maketitle

\section{Introduction}\label{sec:intro}

\sloppy

The nucleus $^{72}$Se sits in the heart of a region of rapidly evolving nuclear shapes and shape coexistence, in which higher mass isotopes show prolate ground states, while the more neutron deficient nuclei are predicted to exhibit rarer oblate deformed ground states \cite{LECOMTE1977123,PhysRevC.18.2801,KAVKA1995177,PhysRevLett.84.4064,Heese1986,Jack72,PhysRevLett.100.102502}.
Predicting the point of this apparent shape transition provides an extremely sensitive test of nuclear models. 
It was recently confirmed by Coulomb excitation that the ground state of $^{72}$Se is dominated by prolate deformation \cite{Jack72}, while the isotopic and isobaric even-even neighbours $^{70}$Se \cite{PhysRevLett.100.102502} and $^{72}$Kr \cite{Wimmer2020,bouchez2003} both show indications of oblate ground states.
However, the situation is complicated by substantial mixing between coexisting configurations and the additional consideration of triaxiality, which has been clearly observed in  $^{72}$Ge \cite{AYANGEAKAA2016254} and $^{76}$Se \cite{Henderson2019}, the presence of which undermines simpler experimental interpretations of deformation. Only with sufficient experimental data to accurately disentangle the mixing of coexisting configurations can we comment on the underlying structure.

Due to quantum mechanical constraints, the non-sphericity of a $0^+$ nuclear state cannot be directly observed.
Deformation can be inferred, either from the $B(E2;0^+\rightarrow2^+)$ reduced transition strength, or by measurement of the diagonal matrix element of the associated $2^+$ state. However, this relies on the model assumption, typically the axially symmetric rotor model. Quadrupole deformation parameters may be determined by applications of model-independent sum rules \cite{SREBRNYCLINE}. Unfortunately these require the measurement of many linking matrix elements, which can in principal be obtained through low-energy Coulomb excitation, but is particularly experimentally challenging when dealing with excited $0^+$ states.
Due to the simplicity of the monopole operator, $\rho^2(E0;0_i^+\rightarrow0_f^+)$ values can be directly equated to the difference in mean square charge radii and the degree of mixing \cite{NewE0Review}, and do not rely on any assumption of axial symmetry.

In this paper we report on new measurements of both the lifetime and branching ratio of the $0_2^+$ state in $^{72}$Se, which in turn yield new values for the $\rho^2(E0)$ between the $0_2^+$ and ground state and the $B(E2)$ from the $0_2^+$ state to the $2_1^+$ state. The new values are consistent with previous measurements and indicate a high degree of state mixing. 

\section{Experiment}

\begin{figure}[!htb] 
    \centering
    \includegraphics[width=0.9\linewidth]{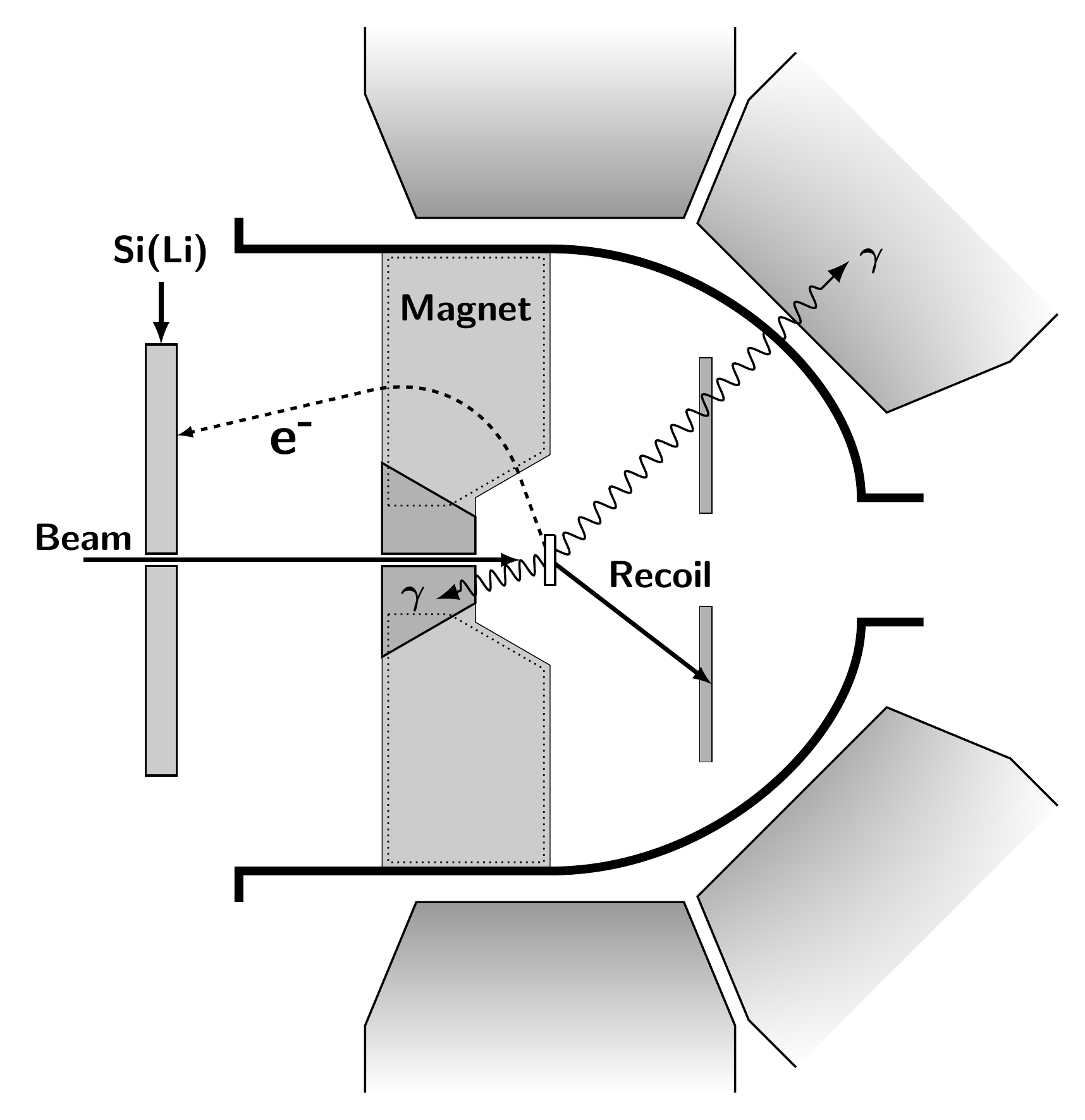}
    \caption{SPICE setup inside the TIGRESS array, with a single downstream recoil detector. The $^{36}$Ar beam enters from the left and passes through openings in both the SPICE Si(Li) detector and photon shield before impinging on the silver backed nat. Calcium target.}
    \label{fig:spicefigure}
\end{figure}

\begin{figure}[!htb] 
    \centering
    \includegraphics[width=0.8\linewidth]{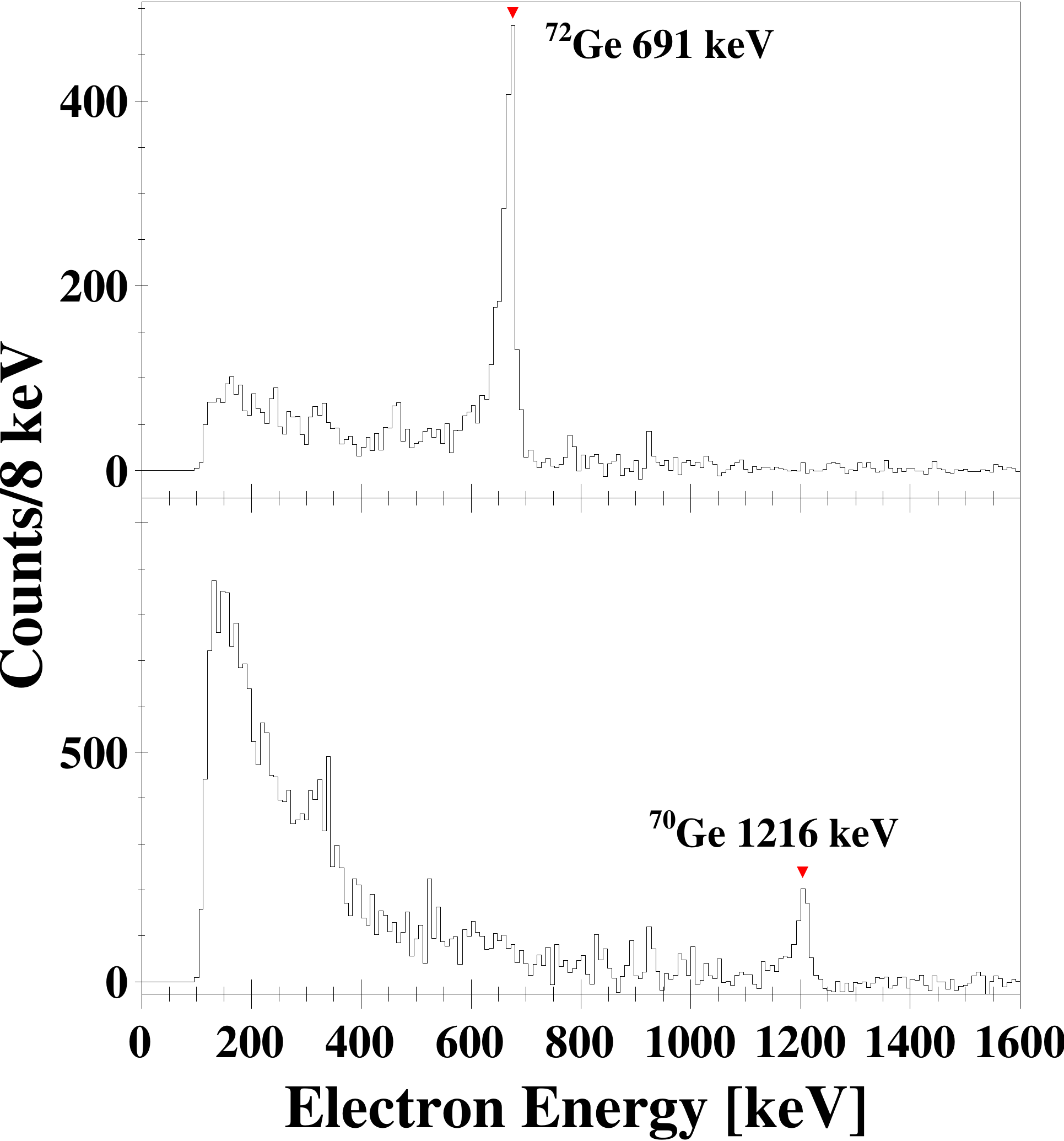}
    \caption{SPICE data showing internal conversion electron peaks associated with $E0$ transition calibration points from $^{70}$Ge (bottom) and $^{72}$Ge (top). Each are selected in coincidence with feeding $\gamma$-ray transitions detected in TIGRESS.}
    \label{fig:E0Ge}
\end{figure}

\begin{figure}[ht!]
  \centering
  \includegraphics[width=0.99\linewidth]{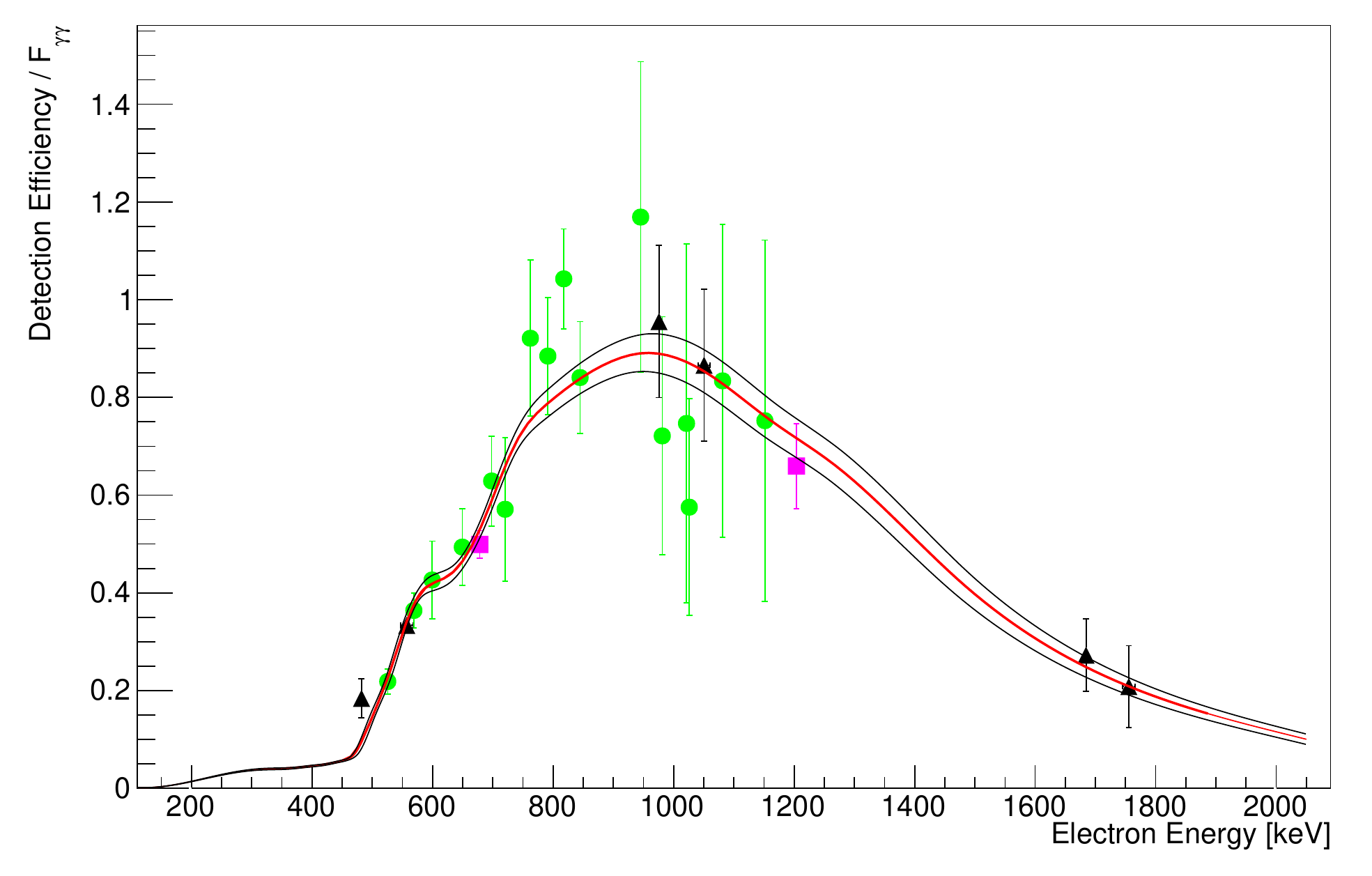}
\caption{Efficiency curve of the SPICE detector (red line) determined from Geant4 simulations \cite{Moukaddam2018}, scaled to in-beam $E2$ and $M1$ data (green circles), $E0$ data (pink squares) and offline $^{207}$Bi data (black triangles). See text for details. One sigma Uncertainty bands resulting from the scaling (black lines) are shown.}
\label{fig:Eff}
\end{figure}

A beam of $^{36}$Ar ions was produced by a microwave ion source \cite{OLIS} and delivered at 120\,MeV (3.33~$A$MeV) by the TRIUMF-ISAC-II linear accelerator chain \cite{ISAC-IIa,ISAC-IIb} to the TIGRESS spectrometer \cite{Hackman14} with an average intensity of 1\,pnA. 
The beam was incident on a target of 0.5\,mg/cm$^2$ nat. Calcium. The target was backed by a 0.2\,mg/cm$^2$ gold ``adhesive'' layer and a 15.7\,mg/cm$^2$ silver beam stopper. The upstream side of the target was sealed by a 0.3\,mg/cm$^2$ layer of gold to prevent oxidisation of the calcium during transfer to the target chamber. Additionally the chamber was flushed with xenon gas prior to target mounting. Despite these precautions, contaminant reactions from the $^{36}$Ar beam on oxygen were observable in the resultant data.

$^{72}$Se nuclei were produced in the fusion-evaporation reaction $^{40}$Ca($^{36}$Ar,4p)$^{72}$Se.
Recoiling selenium nuclei were stopped in the silver backing with a stopping time of approximately 0.2 ps.
$^{72}$Se was also populated indirectly via beta decay of $^{72}$Br ($t_{1/2}$=78.6~s \cite{ABRIOLA20101}) produced in the $^{40}$Ca($^{36}$Ar,3pn)$^{72}$Br reaction, with a cross-section approximately 10\% that of the former reaction (PACE4).
Evaporated protons were detected by a Micron S3 silicon detector of 140\,$\mu$m thickness located downstream of the target to aid in selection of the reaction channel of interest.

Twelve of the TIGRESS high-purity germanium (HPGe) clover detectors were positioned around the target location to detect $\gamma$ rays. Four clovers were located at 45$^{\circ}$ with respect to the beam axis and eight at 90$^{\circ}$. Each clover was Compton suppressed and positioned at a target-to-detector distance of 14.5\,cm in order to optimise the peak-to-total configuration of the TIGRESS spectrometer \cite{Hackman14}.

The Spectrometer for Internal Conversion Electrons (SPICE) \cite{Moukaddam2018,Ketelhut2014,refId0,GarnsworthyEPJ} was used to detect internal conversion electrons. SPICE utilises a 6.1\,mm thick lithium-drifted silicon [Si(Li)] detector located upstream from the reaction target, and shielded from direct sight by a photon shield. A magnetic lens formed of rare-earth permanent magnets collects and directs internal conversion electrons around the photon shield to the Si(Li) detector. 

The detector signals were processed by the TIGRESS data acquisition system \cite{Martin08}. Data were recorded to disk for every event in which a Si(Li) trigger was detected. Additional events were recorded for a $\gamma-\gamma$ trigger, these events were down-scaled by 4 or 8, manually selected dependant on the data rate.

The absolute $\gamma$-ray efficiency of TIGRESS was obtained from standard radioactive sources of $^{133}$Ba, $^{152}$Eu, $^{207}$Bi, and $^{56}$Co.
\newline
\newline
\textbf{SPICE Efficiency}

As fusion-evaporation products were implanted into a thick target backing, additional straggling of electrons, especially at low energies, reduced the efficiency compared to that determined from using ``open'' offline radioactive sources, which have minimal scattering material covering the activity.

The relative in-beam electron detection efficiency curve of SPICE was obtained from a detailed Geant4 simulation of the setup \cite{Ketelhut2014,Smallcombe2018}.
This curve was normalized to measurements of internal conversion coefficients (ICCs) for known transitions for which the calculated ICCs are reliable \cite{Kibedi2007,Kibedi2008}. It was possible to use 24 $\gamma$ ray gated known $E2$ and $M1$ transitions from  in-beam products $^{107,109}$Ag, $^{72}$Se, $^{73}$Br and $^{70}$Ge.
Additional data points were determined from the ratio of $0_2^+{\rightarrow}0_1^+$ $E0$ ICEs to other ($E2$) transitions. In the case of $^{70}$Ge a competing $0_2^+{\rightarrow}2_1^+$ transition allowed an easy comparison to the well measured branching ratio. For $^{72}$Ge the $0_2^+$ is the lowest excited state with no competing branch. However, $^{72}$Ge was entirely populated in beta decay, having two additional neutrons with respect to the compound nucleus $^{76}$Sr\footnote{As natural calcium was used in the target the 0.647\% $^{42}$Ca and 2.09\% $^{44}$Ca would allow the $^{42}$Ca($^{36}$Ar,6p)$^{72}$Ge and $^{44}$Ca($^{36}$Ar,4p$\alpha$)$^{72}$Ge population channels, however temporal gating with the beam-RF confirmed these prompt contributions are imperceptible at the sensitivity level of this work.}. Consequently the ratio of the $0_2^+{\rightarrow}0_1^+$ $E0$ peak, which has no competing branch, can be taken with respect to the $2_1^+{\rightarrow}0_1^+$ transition using the well measured decay scheme of $^{72}$As \cite{ABRIOLA20101}. The associated electron peaks for the $^{70,72}$Ge $E0$ calibration points are shown in Figure~\ref{fig:E0Ge}

Additional data were obtained from an offline $^{207}$Bi source measurement, the source data points were adjusted for the change in position between the source and target, and the depth of implantation, using a Geant4 simulation. A systematic uncertainty was included in each $^{207}$Bi point, equal to 30\% of the adjustment, this was found to be sufficient to ensure a reduced chi-squared of less than 1 between the $^{207}$Bi points and the final efficiency curve.
The validity of the simulation was further qualified by comparing the spectra from an un-shielded $^{207}$Bi source and that obtained when placing an aluminium foil of known thickness obscuring the source as a surrogate for implantation.

Finally the measured points were spanned by the Geant4 simulations to give the resultant efficiency curve and uncertainty shown in Figure \ref{fig:Eff}. The in-beam efficiency calculated is proportional to $F_{\gamma\gamma}$, the down scaling factor of the $\gamma\gamma$ coincidence data. Due to changes to the downscaling factor of the $\gamma\gamma$ trigger condition during the data collection $F_{\gamma\gamma}$ cannot be accurately determined, however as the factor cancels out in experimental measurements the value is not required.

\section{Results}

\begin{figure*}[ht!]
  \centering
  \includegraphics[width=0.8\linewidth]{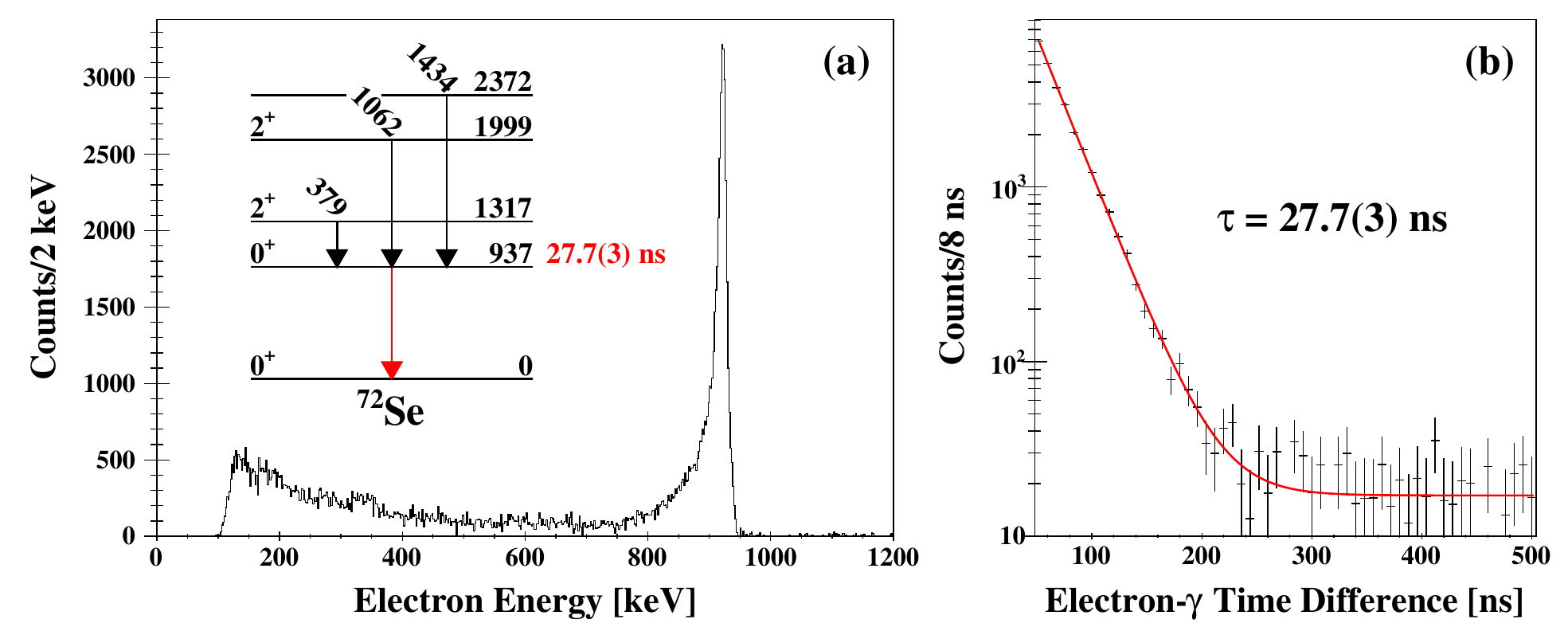}
\caption{(a) Electron spectrum produced from a sum of three $\gamma$-ray gates, 379, 1062 and 1434 keV, showing the $0_2^+{\rightarrow}0_1^+$ $E0$ transition of $^{72}$Se cleanly isolated. (b) Electron-$\gamma$ time difference spectrum between transitions populating and depopulating the $0^+_2$ state in $^{72}$Se, produced from a sum of the same three populating $\gamma$-ray gates as (a). A fit to the data yields a lifetime measurement of 27.7(3)~ns.
}
\label{fig:Life}
\end{figure*}

\begin{table}
  \begin{center}  
  \caption{Gamma-ray branching ratios for the $^{72}$Se 1317~keV $2_2^+$ state, B$_{\gamma,expt}$, from the present work, alongside McCutchan \textit{et al.} B$_\text{McC}$ \cite{McCutchan}, Mukherjee \textit{et al.} B$_\text{Muk}$ \cite{New72paper}, and the 2009 evaluation values B$_\text{lit}$ \cite{ABRIOLA20101}.}
  \label{table:twobranch}
    \begin{tabular}{ccccc} 
     \hline
     \hline
     {E$_{\gamma,expt}$} (keV)& {B$_{\gamma,expt}$}& {B$_\text{McC}$} & {B$_\text{Muk}$}  & {B$_\text{lit}$}\\
     \hline 
     1316.5(4) & 100(3)& 100(5) & 100(10) & 100(6)\\
     454.6(5) & 79(2) & 77(4) & 167(20) & 76(5)\\
     379.3(2) & 16(2) & 15(1) & 22(3) & 35(2)\\
     \hline 
    \end{tabular}
  \end{center}
\end{table}

\begin{figure}[ht!]
  \centering
  \includegraphics[width=0.9\linewidth]{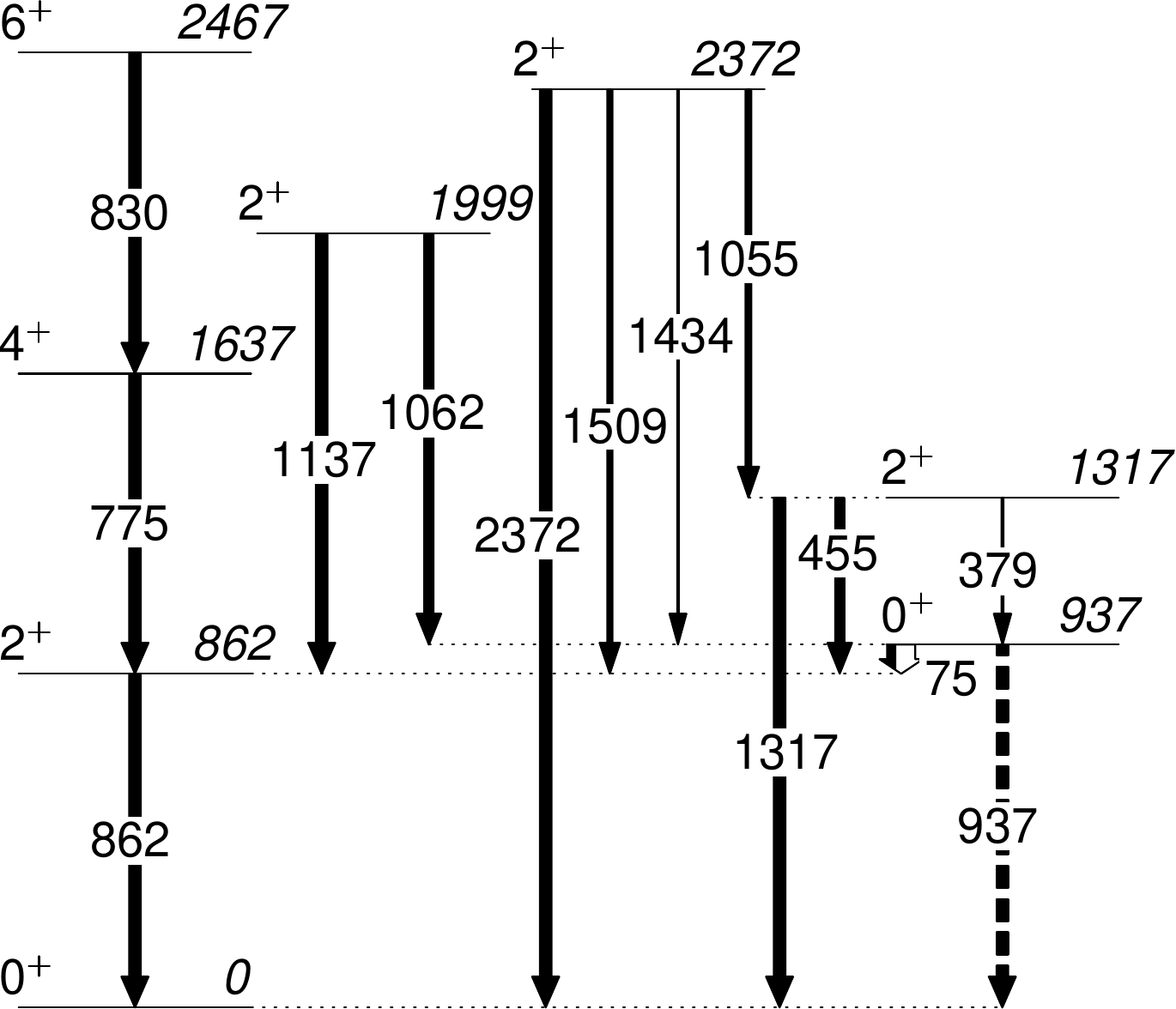}
\caption{Partial level scheme of $^{72}$Se highlighting the lowest lying positive parity states of interest. Relative arrow widths indicate branching ratios for individual state and the proportion of white indicates the level of internal conversion. The dashed line denotes an $E0$ transition. Level and transition energies are labelled in keV.}
\label{fig:partscheme}
\end{figure}

To determine precise values for the $\rho^2(E0)$ and $B(E2)$ strengths of transition depopulation the $0_2^+$ state in $^{72}$Se, both a precise measurement of the state lifetime and of the transition branching ratio was performed.

Energy gated time difference spectra were produced for $\gamma$-ray transitions populating and internal conversion electrons depopulating the 937~keV $0_2^+$  state. The internal conversion electron peak corresponding to the $E0$ transition to the ground state, shown in Figure~\ref{fig:Life}, provided an unambiguous gate leading to a clean and distinct time difference lifetime spectrum. The internal conversion peaks of K-, L- and higher-shell electrons for selenium cannot be resolved within the energy resolution of SPICE obtained in this beam time and a single energy gate was used.

Time differences were determined using the digital data acquisition system. Event times for $\gamma$ rays detected in TIGRESS were determined online within the constant fraction discrimination algorithm of the TIG10 digitizer firmware. The $\gamma$-ray timing distribution suffered from significant tailing due to charge collection times. Waveforms from SPICE were recorded at 100~MHz and fit offline to determine accurate times for electron detection. This fitting procedure for SPICE ensured there was no tailing due to electron time measurement and subsequently the electron-$\gamma$ timing distribution remained Gaussian on the side of the timing distribution on which the lifetime was to be measured. The combined timing resolution for electron-$\gamma$ coincidence events was determined to be $\sigma_t\sim$14 ns. 

A sum of three different $0_2^+$ state feeding $\gamma$-ray gates (379, 1062, 1434~keV) was used to produce the spectra shown in Figure \ref{fig:Life}. The shape of the detector response for the HPGe-Si(Li) time difference was determined experimentally from prompt ($\sim$ps lifetime) transitions and incorporated into the fit.

To ensure the contribution of the energy-dependant $\gamma$-ray time distributions were well understood, time spectra were simulated for the individual $\gamma$-ray gates, based on the experimentally determined prompt time spectra and a variable lifetime. These simulated spectra were fit to the individually gated data and minimised to extract a lifetime. The results for each individual gate agreed with that of the combined fit.

A value of $\tau=$27.7\,(3)\,ns was determined for the lifetime of the $0^{+}_{2}$ state. This agrees with the previous measurement \cite{PhysRevC.9.948} of $\tau=$27.8\,(6)\,ns. However, the value of $\tau=$22.8\,(14)\,ns reported by \cite{PhysRevLett.32.239} is significantly discrepant and so the Normalised Residual Method (NRM) was used to determine a weighted average  \cite{NMR,EvalGuide}. A value of $\tau(0_2^+)=$27.7\,(4)\,ns is adopted.

In order to determine the $E0/E2$ branching ratio for depopulation of the 937~keV state, $\gamma\gamma$ and electron-$\gamma$ data were selected by coincidence with the 379\;keV and 1062\;keV $\gamma$-ray feeding transitions.
As the competing $0_2^+{\rightarrow}2_1^+$ $E2$ transition is only 75.06(20)\,keV \cite{McCutchan} it could not be easily resolved in the data and the efficiency of the TIGRESS array at this energy is poorly constrained. Instead the area of the subsequent 862\,keV $2_1^+{\rightarrow}0_1^+$ transition was used, which due to direct feeding can be equated to the intensity of the depopulating $E2$ branch. Branch intensities are given by
\begin{equation}
    I(E2)=\frac{N_\gamma\left(1+\alpha\right)}{\epsilon_{\gamma}\epsilon_{\gamma_{g}}F_{\gamma\gamma}\Delta t_{\gamma\gamma}},
\end{equation}
where $N_\gamma$, $\alpha$ and $\epsilon_{\gamma}$ are the measured counts, conversion coefficient and $\gamma$-ray detection efficiency for the 862\,keV transition, and by
\begin{equation}
    I(E0)=\frac{N_e\Omega_{Tot}}{\epsilon_{e}\epsilon_{\gamma_{g}}F_{\gamma\gamma}\Delta t_{e\gamma}\Omega_{e}},
\end{equation}
where $N_e$, $\Omega$ and $\epsilon_{e}$ are the measured counts, electronic factors and electron detection efficiency for the $E0$ transitions.
As previously discussed the experimental electron efficiency $\epsilon_{e}=\epsilon_{Abs}/F_{\gamma\gamma}$.
The efficiency for the gating transition, $\epsilon_{\gamma_{g}}$ and the $\gamma\gamma$ down-scaling factor, $F_{\gamma\gamma}$, need not be evaluated as these cancel when the ratio is taken with $I(E0)$. The coincidence time fraction $\Delta t$ must be calculated as it is different for $\gamma\gamma$ and electron-$\gamma$ coincidences.

Values of $I_{E0}/I_{E2}=0.47(6)$ and $q^2_K(E0/E2)=0.71(9)$. were determined, which may be related by \cite{PhysRev.109.1299}
\begin{equation}
   q^2_K(E0/E2)=\frac{I(E0)}{I(E2)}\times\frac{\Omega_{K}(1+\alpha)}{\Omega_{Tot}\alpha_K},
\end{equation} 
where the conversion coefficients are for the $0_2^+{\rightarrow}2_1^+$ transition. Values of $\alpha=2.42(4)$ and $\alpha_K=2.042(35)$ are taken from BrIccS v2.3d \cite{Kibedi2008}. Electronic factors $\Omega_{Tot}=3.46(17)\times10^{8}$ and $\Omega_{K}=3.09(15)\times10^{8}$ are from the latest mass-corrected relativistic calculations \cite{jackson,jackson2}.

Taking a weighted average of $I_{E0}/I_{E2}$ with previously reported values 0.38(20) \cite{PhysRevC.12.1360}, 0.41(8) \cite{McCutchan} and 0.32(4) (converted from measured $q^2_K$) \cite{ithemba,Priv_Comm} yields a final value of $I_{E0}/I_{E2}=0.38(4)$, with a reduced $\chi^2$=1.70 confirming reasonable agreement of all data.\footnote{As the reduced $\chi^2$ value (1.61) of the weighted average is less than the 95\% critical value (2.60), the data are not deemed to be discrepant \cite{CHECHEV2000601}, however ``external error'', inflated by $\sqrt{\chi^2_{r}}$, is used.}

Combining the weighted branching ratio with the lifetime determined above, an $E0$ strength $\rho^2(E0;0_2^+{\rightarrow}0_1^+)=29(3)$ milliunits is extracted, using the relation
\begin{equation}
\rho^2(E0)=
\frac{1}{
\tau\,
\left(
1+\frac{1}{I_{E0}/I_{E2}}\right) \Omega_{Tot}
}\,.
\end{equation}

This updates the previous value of 30(5) milliunits \cite{NewE0Review} with an increase in precision.
A value of $B(E2;0_2^+{\rightarrow}2_1^+)=148(5)$\,W.u. is also determined, which is in agreement with previous values.

In addition, branching ratios for $\gamma$-rays depopulating the 1317~keV $2_2^+$ state in $^{72}$Se were measured. The values are shown in Table~\ref{table:twobranch}. These are consistent with those reported by McCutchan \textit{et al.} \cite{McCutchan} confirming a smaller feeding of the $0_2^+$ state than determined in previous literature evaluation \cite{ABRIOLA20101}. In appears that in the evaluation the 455 keV branch intensity is taken from $^{72}$Br $\varepsilon$ decay, while the 379 keV branch is taken from in-beam reactions, as the 379 keV $\gamma$ ray was multiply placed in the decay data. However, the 379 keV intensity from $\varepsilon$ decay ($I_\gamma\leq21$) agrees reasonably with the value reported here. A recent measurements by Mukherjee \textit{et al.} \cite{New72paper} reports discrepant branching ratios, which are excluded from our discussion. 

\section{Discussion}

\begin{table*}
  \begin{center}  
  \caption{Results of two state mixing calculations for $0_1^+$	and	$2_1^+$ states in $^{72}$Se comparing experimental and calculated spectroscopic quadrupole moment $Q_s(2_1^+)$ and electric monopole transition strength $\rho^2(E0;0_2^+\rightarrow0_1^+)$. The $0^+$ and $2^+$ mixing matrix elements $\braket{h_0}$ and $\braket{h_2}$ are given alongside the degree of state mixing, given by the mixing coefficient $b$ squared.}
  \label{table:mixresults}
    \begin{tabular}{rcccccccc} 
	&	$Q_s(2_1^+)$ 	&	$\rho^2(E0)$	&	{\multirow{2}{*}{$\beta_a$}}	&	{\multirow{2}{*}{$\beta_b$}}	&	$\braket{h_0}$	&	$\braket{h_2}$	&	$0_1^+$	&	$2_1^+$	\\
	&	eb	&	$\times10^{-3}$	&		&		&	keV	&	keV	&	$b_0^2$	&	$b_2^2$	\\
  \hline	  \hline																
Experiment	&	-0.57(31)$^{\dagger}$	&	29(3)	&		&		&		&		&		&		\\
  \hline																	
Spherical-prolate$^{\dagger\dagger}$&	-0.29	&	39	&	0	&	0.31	&	231	&	227	&	0.07	&	0.47	\\
  \hline																	
Oblate-Prolate	&	-0.13	&	28	&	-0.01	&	0.21	&	473	&	221	&	0.40	&	0.35	\\
Oblate-Oblate	&	0.02	&	31	&	-0.01	&	-0.22	&	420	&	30	&	0.27	&	0.01	\\
Prolate-Oblate	&	-0.57	&	29	&	0.3	&	-0.21	&	428	&	29	&	0.30	&	0.04	\\
Prolate-Prolate	&	-0.55	&	29	&	0.28	&	0.36	&	410	&	32	&	0.25	&	0.01	\\
    \hline
    \end{tabular}
  \end{center}
    \footnotesize\vspace{-6pt}
\setlength\tabcolsep{1.5pt}
\begin{tabular}{rl}
\hspace{2.6cm} $^{\dagger}$ & Experimental quadrupole moment is taken from Ref. \cite{Jack72}.\\
$^{\dagger\dagger}$ & Spherical-prolate results are derived from Ref. \cite{McCutchan}.
\end{tabular} 
\end{table*}

\begin{figure*}[ht!]
  \centering
  \includegraphics[width=0.8\linewidth]{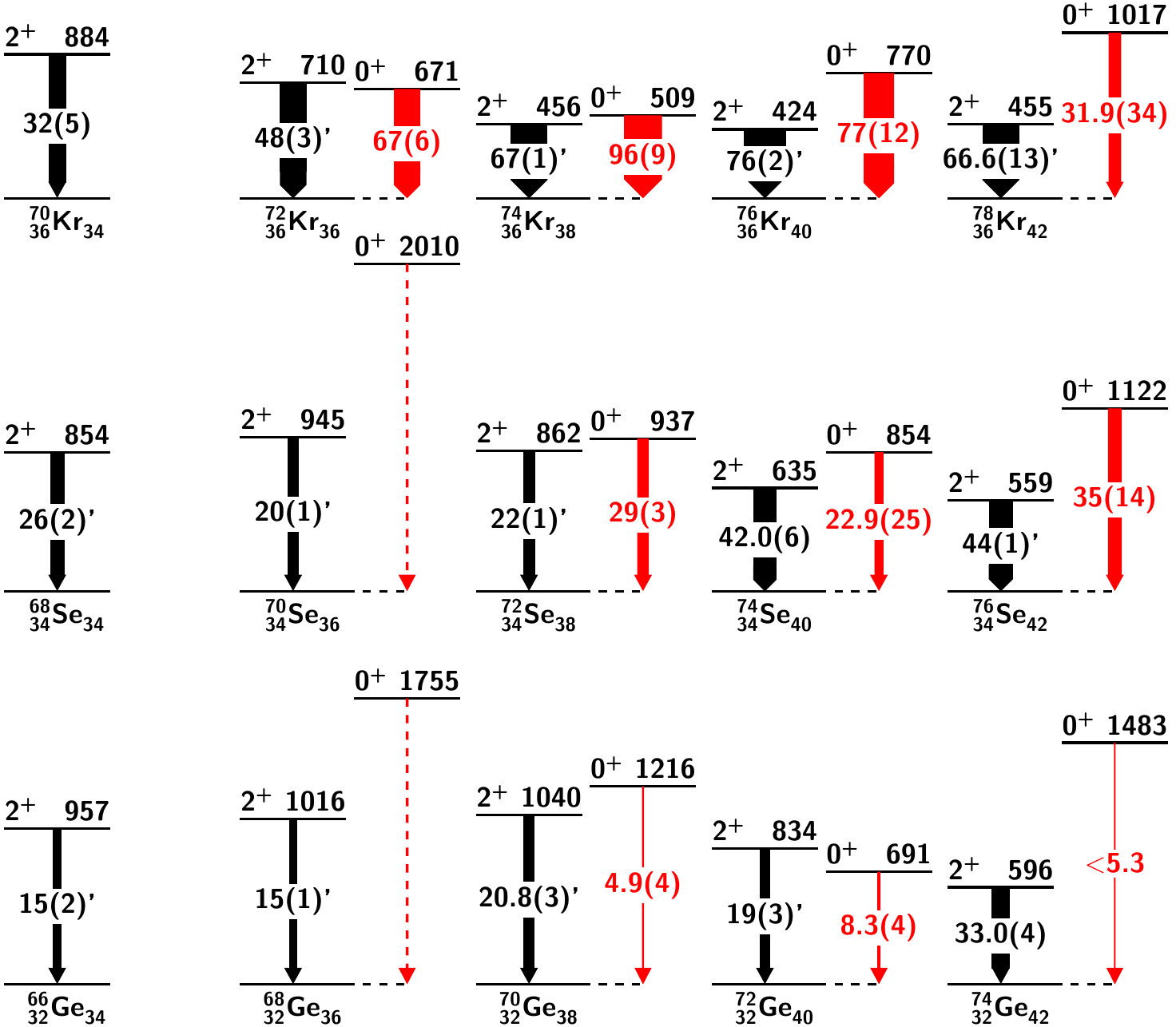}
\caption[]{Figure showing the lowest known excited $2^+$ and $0^+$ states in Kr, Se and Ge isotopes. Known, $B(E2)$ transition strengths in W.u. and $\rho^2(E0)$ values in milliunits are given and are indicated by  the relative line width of transition arrows. A dashed line denotes a possible $E0$ transition which has not been directly observed.
 $\rho^2(E0)$ values are from Ref \cite{NewE0Review}, except $^{72}$Se. $B(E2)$ values with apostrophes are weighted averages between published data tables \cite{ABRIOLA20101,BROWNE20101093,MCCUTCHAN20121735,GURDAL20161,SINGH20061923,FARHAN20091917,SINGH1984233} and more recent data \cite{AYANGEAKAA2016254,BECKER2006107,PhysRevC.75.054313,PhysRevLett.96.189901,Gorgen2005,Henderson2019,iwasaki2014,Leske2006,PhysRevLett.100.102502,LUNDERBERG201830,Lutt2021,MORSE2018198,NICHOLS201452,Obert2009,Palit2001,Wimmer2020,Wimmer2021} (not evaluated values).}
\label{fig:NZdiag}
\end{figure*}

Previous studies of $^{72}$Se \cite{McCutchan,lieb} suggested that the low-lying states of the nucleus can be partially described if one assumes there is a crossing of a rotational band built on a prolate $0_2^+$ state, with states of a near-spherical vibrator band built on the ground state, in which mixing is limited to the $2^+$ and $4^+$ states. The remarkable agreement in the energy of the $0_2^+$ with the yrast $J>4$ states when fit with a polynomial of $I(I+1)$ reported in Ref. \cite{lieb} would seem to agree with this characterisation. However, this assignment is based on the assumption of minimal mixing between the $0^+$ states, which is not supported by the moderate $\rho^2(E0)$ strength. When one introduces mixing of the $0^+$ states, the unperturbed energy of the initially pure $0^+_{rotor}$ will be reduced in energy, relative to the observed $0_2^+$ state, and have poorer agreement with the described $I(I+1)$ fit.

In the previous works \cite{McCutchan,lieb}, small $0^+$ mixing was suggested to explain the over-prediction of the $2_2^+\rightarrow0_1^+$ strength. However, the authors wrongly conclude that this mixing would produce destructive interference in the $2_2^+\rightarrow0_1^+$ transition and constructive interference in the $2_1^+\rightarrow0_2^+$ transition. In the model used, the interference terms in these transitions would be of the same sign.

While $B(E2)$ values have received much attention, in the present work the somewhat neglected interpretation of $\rho^2(E0)$ strength is considered, along with the recently obtained experimental spectroscopic quadrupole moment $Q_s(2_1^+)$ \cite{Jack72}.
Table \ref{table:mixresults} shows results of a two state mixing calculation. This includes the spherical-prolate results from Ref. \cite{McCutchan} alongside new calculations for coexisting rotor bands, which use the same basis as \cite{McCutchan,lieb}, taken from \cite{Dickmann}.
The ratio of unperturbed level spacing between the intrinsic rotor bands was given by the rigid-body moment of inertia as
\begin{equation}
    \frac{E(2_a^+)-E(0_a^+)}{E(2_b^+)-E(0_b^+)}
    =
    \frac{1+0.315\beta_b+0.249\beta_b^2} 
    {1+0.315\beta_a+0.249\beta_a^2} 
\end{equation}
in which a negative value for $\beta_a/\beta_b$ is adopted for an oblate shape of band $a/b$.
With the exception of the imposed mixing matrix elements, the two structures are assumed to be independent and to have negligible inter-band matrix elements. Only mixing of $0^+$ and $2^+$ states was considered, constraints were not imposed based on high-lying yrast states.

Electric monopole transition strengths between mixed states are calculated as \cite{Wood1999} 
\begin{equation}
    \rho^2(E0)=a_0^2b_0^2\left(\frac{3}{4\pi}\right)^2Z^2\left(\beta_a^2-\beta_b^2\right)^2
    \label{eq:E0}
\end{equation}
where $a_0$ and $b_0$ are the mixing coefficients of the $0^+$ states. 
The intrinsic quadrupole moments $Q_0$ of the unmixed bands are taken as a function of $\beta_2$ under a sharp-surface approximation using
\begin{equation}
Q_0=\frac{3}{\sqrt{5\pi}}ZR_0^{2}\beta
\left(1+0.16\beta\right)
\end{equation}
The mixed state quadrupole moments were taken as
\begin{equation}
    Q_0=a^2Q_a+b^2Q_b
\end{equation}
under the stated independence assumption i.e. $\bra{2_b}\hat{Q_2^0}\ket{2_a}$$\simeq$0.
This was related to the observed spectroscopic quadrupole moment by the usual axial-rotor expression
\begin{equation}
    Q_s=\frac{3K^2-I(I+1)}{(I+1)(2I+3)}Q_0
\end{equation}

Within the limitations of the twin-rotor model,
in order to reproduce the significantly different $E(2^+)-E(0^+)$ energy spacing of the observed bands, a significantly greater mixing of the $0^+$ than $2^+$ states is required. A prolate-oblate shape coexistence description, with equal $0^+$ and $2^+$ mixing matrix elements, can only be achieved with unphysically large deformation parameters ($\beta>1$).
Reproducing the observed $\rho^2(E0)$ then strongly limits the magnitude difference of deformation of the two bands, per Equation \ref{eq:E0}.
Consequently it is observed that, in order to reproduce the negative $Q_{s}(2_1^+)_{,expt.}$, the $0_a^+$ band must be prolate in character, as the deformation of $0_b^+$ band cannot differ sufficiently to change the sign of the observed $2_1^+$ state deformation.
Subsequently, when the $0_a^+$ state is constrained to negative $\beta$ values, it tends towards spherical shape.
The calculations are relatively insensitive to the intrinsic deformation of the $0_b^+$ state.
The measured $\rho^2(E0)$ value effectively places a lower limit on the degree of mixing.
Combining this with the experimental quadrupole moment all but rules out an oblate ground state, within the limitations of the chosen model basis. 
However, the disproportionately large $\braket{h_0}$ value determined, and poor agreement with higher-lying yrast states, show further factors must be included to describe the structure of $^{72}$Se.

Given the presence of triaxiality recently identified in $^{76}$Se \cite{Henderson2019} and in neighbouring $^{72}$Ge \cite{AYANGEAKAA2016254}, a degree of triaxiality in the ground state of $^{72}$Se may be expected. Such triaxiality may be suppressed in high-spin yrast states through centrifugal effects. Current models do not explain the behaviour of the $2_3^+$ at 1999\,keV \cite{PhysRevC.95.064310}, which might be considered to be second $2^+$ state of a triaxial configuration. However, as shown in Fig.~\ref{fig:partscheme}, the state at 1999\,keV feeds both the $0_2^+$ and $2_1^+$ with approximately equal strength, but not the ground state, to which one would expect a branch to be 20 times larger based on energy weighting.
Mukherjee \textit{et al.} \cite{New72paper} do report observation of a small $2_3^+\rightarrow0_1^+$ branch, however they did not report the 1062\,keV transition, previously assigned as $2_3^+\rightarrow0_2^+$, which was clearly observed in the present work as directly feeding $0_2^+$ state.

One may also use model independent sum rules to extract quarupole shape invariants $Q^2$ and $cos(3\delta)$ \cite{PhysRevLett.28.249}. The complete summation should be over all $E2$ matrix elements of a nucleus, but it has been demonstrated that an approximate value can be determined from the first terms of the summation \cite{PhysRevC.102.054306}. For the ground state of $^{72}$Se the first two terms can be calculated, yielding values of $Q^2\approx0.21(2)e^2b^2$ and $cos(3\delta)\approx0.3(2)$, in which the uncertainties are from experimental values and do not represent the effect of the curtailed summation. This corresponds to $\delta\approx24\si{\degree}$, indicating a significant degree of triaxiality.

While Table \ref{table:mixresults} shows that a spherical ground state can reproduce the observed $Q_s$ and $\rho^2(E0)$ values moderately well, without resorting to the extreme matrix elements observed in the twin-rotor calculations, this interpretation does not match with the apparent triaxiality.

Figure \ref{fig:NZdiag} shows the first exited $2^+$ and $0^+$ states of even-even isotopes for $32{\leq}Z{\leq}36$ and $34{\leq}N{\leq}42$. 
While many low-lying $0^+$ states are present, the smooth parabolic trajectories associated with shape-coexistence in higher-mass nuclei are not clearly observed. The krypton isotopes show hints of this trend, but the selenium chain is somewhat flatter, with the exception of $^{70}$Se.
The measured $\gamma$-ray branching ratio of the $2_2^+$ state in $^{72}$Se may indicate a larger degree of mixing between the two bands than previously assumed, however the measured $E0$ transition strength for $^{72}$Se, being less than half of the 71(6) milliunits of $^{72}$Kr, would seem to support a description of Se isotopes with less significant $0^+$ state mixing than neighbouring Kr isotopes.
However what has been shown above is that the degree of mixing can be just as
great in the selenium isotopes if the magnitude difference of the co-existing shapes is smaller. To obtain a more complete picture of the structure of these nuclei requires the measurement of both $E0$ transition strength and quadrupole moments of other low-lying states.

Simple two-state mixing models fail to describe $^{72}$Se adequately and there is significant evidence that any description based on mixing of distinct independent structures will not adequately capture the nature of the nucleus. Robust mean-field calculations which can construct states in a shared potential of multiple minima are called for.

\section{Conclusion}

Utilising the SPICE and TIGRESS arrays, an independent measurement of the branching ratios and lifetime of the $0_2^+$ state in $^{72}$Se was performed by direct observation of internal conversion electron and $\gamma$ rays. Combining these experimental observables a new evaluation of the $E0$ and $E2$ strength between the $0_2^+$ state and the ground state was performed. Values of $\rho^2(E0;0_2^+{\rightarrow}0_1^+)=29(3)$ milliunits and $B(E2;0_2^+{\rightarrow}2_1^+)=148(5)$\,W.u. were determined, improving on the precision of the previous values.
Mixing axial-rotor model calculations were performed to explore the significance of the observed $\rho^2(E0)$ in combinations with the recently measured spectroscopic quadrupole moment.
Within the confines of the model it was demonstrated that the deformation of the ground state band should be prolate in nature and that mixing between intrinsic $0^+$ states in $^{72}$Se may be significant and should not be neglected in calculations. However, it is concluded that independent two-state mixing does not describe the nucleus well overall.

\begin{acknowledgments}
We would like to thank R. Herzberg for helpful discussions. Thanks also to the beam delivery and technical staff of the TRIUMF-ISAC facility. The SPICE infrastructure was funded by the Canada Foundation for Innovation and the Ontario Ministry of Research and Innovation. TRIUMF receives funding through a contribution agreement through the National Research Council Canada. C.E.S. acknowledges support from the Canada Research Chairs program.
This work was supported in part by the Natural Sciences and Engineering Research Council of Canada (NSERC), by the U.S. National Science Foundation Grant No. 1606890, and by the Science and Technology Facilities Council (STFC) Grant No. ST/P003885/1 and ST/V001035/1.
We would like to thank M. Pearson for consultation during the setup and operation of SPICE.
\end{acknowledgments}

\bibliography{72SeSPICE_PRCAndOthers}

\end{document}